\documentclass[cits]{PoS}
\graphicspath{{figures/}}

\title{Vacuum structure and Casimir scaling in Yang--Mills theories\addtocounter{footnote}{1}\thanks{This research was supported in part by the Slovak Grant Agency for Science, Project VEGA No.\ 2/0070/09, by ERDF OP R\&D, Project CE QUTE ITMS 262401022, and via CE SAS QUTE.}}

\ShortTitle{Vacuum structure and Casimir scaling in Yang--Mills theories}

\author{\addtocounter{footnote}{-2}\speaker{\v{S}tefan Olejn{\'\i}k}\\
        Institute of Physics, Slovak Academy of Sciences, SK--845 11 Bratislava, Slovakia\\
        E-mail: \email{stefan.olejnik@savba.sk}}

\abstract{The vacuum of Yang--Mills theories can be imagined as a magnetically disordered medium with domain structure, with color magnetic flux in each domain quantized
in units corresponding to the gauge group center. This model leads to the prediction of Casimir scaling, i.e. the proportionality of string tensions of potentials (at intermediate distances) between color sources from higher-representations to eigenvalues of the quadratic Casimir operator. I present evidence for Casimir scaling in \gtwo\ lattice gauge theory. I also discuss support for some ingredients of the model from the recently conjectured form of the Yang--Mills ground-state wave-functional.}

\FullConference{International Workshop on QCD Green's Functions, Confinement and Phenomenology\\
		 September 7--11, 2009\\
		 ECT Trento, Italy}

\newcommand{\Olejnik}{{\v{S}}.~Olejn{\'\i}k}
\newcommand{\gtwo}{$\mathrm{G}_2$}
\newcommand{\arXivold}[2]{\href{http://www.arxiv.org/abs/#1/#2}{\footnotesize\tt arXiv:#1/#2}} 
\newcommand{\arXivnew}[2]{\href{http://www.arxiv.org/abs/#2}{\footnotesize\tt arXiv:#2 [#1]}} 

\begin{document}

\section{Introduction: the Casimir scaling hypothesis}\label{intro}
\begin{figure}[b]
\centerline{\includegraphics[width=0.6\textwidth]{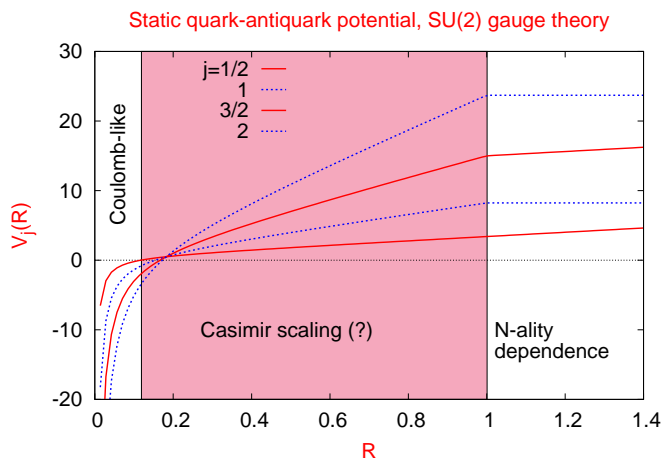}}
\caption{Sketch of the expected behavior of static potentials in SU(2) gauge theory.}
\label{potential}
\end{figure}

The explanation of how quark and gluon confinement arises in non-Abelian gauge theories remains a challenging problem. To make the problem simpler, one can leave out dynamical quark degrees of freedom and ask e.g.\ about the behavior of potentials between static color sources from different representations of the gauge group, but even this question does not have a fully satisfactory answer. The potentials are expected to behave differently in three regions of quark-antiquark distances (the situation is sketched in Fig.\ \ref{potential} for the case of SU(2)). At short distances the interaction is governed by perturbative one-gluon exchange and the potentials are Coulomb-like. At asymptotic distances, the color charges of higher-representation static sources can be screened by an appropriate number of gluons; in SU($N$) gauge theory, potentials of all zero $N$-ality representations become asymptotically flat, while asymptotic string tensions for representations with nonzero $N$-ality will be the same as for the lowest representation with the same $N$-ality. Most interesting is the region of intermediate distances, from the onset of confinement to the onset of screening. There string tensions depend on color-charge representations in a more intricate way: that dependence carries an imprint of the underlying mechanism of confinement.

It was argued that intermediate string tensions of potentials between higher-representation quarks and antiquarks should be proportional to the quadratic Casimir. This \textit{Casimir scaling hypothesis} \cite{DelDebbio:1995gc} can be supported e.g.\ by large-$N$ factorization \cite{Greensite:1982be} or dimensional reduction arguments \cite{Greensite:1979yn,Olesen:1981zp,Ambjorn:1984mb}. Regardless of whether one is convinced by these arguments or not, Casimir scaling is an experimental fact: it was observed convincingly in numerical simulations of both SU(2) and SU(3) lattice gauge theories (most recently in Refs. \cite{Piccioni:2005un,Deldar:1999vi,Bali:2000un}).

Any viable model of the confinement mechanism should be able to explain both Casimir scaling at intermediate and $N$-ality dependence at large distances. It is not easy at all to understand these effects in terms of vacuum fluctuations which dominate the QCD functional integral \cite{Shevchenko:2003ks}.
      
In the present contribution I will first discuss a simple model of how (approximate) Casimir scaling and $N$-ality dependence arise within the center-vortex picture of color confinement, and then extend the model to non-Abelian theories with trivial center (Section \ref{model}). Its specific prediction is Casimir scaling for potentials in (center-trivial) \gtwo\ gauge theory. I will report results of numerical simulations that confirm the prediction (Section \ref{g2_cs}). Finally, I will outline a recent suggestion for an approximate wave-functional of the Yang--Mills theory that incorporates (some) elements of the proposed vacuum model (Section \ref{wf}). Section \ref{conclusions} summarizes conclusions.

\section{A simple model of Casimir scaling and color screening}\label{model}
$N$-ality (or ``representation class'' in mathematical language) is related to the transformation properties of SU($N$)-group representations with respect to the group center, Z${}_N$. The $N$-ality dependence of string tensions of different-representation potentials at asymptotic distances can be easily understood in terms of the picture of the QCD vacuum as a center-vortex condensate, where the dominant field configurations are directly related to elements of the group center \cite{tHooft:1978}. The asymptotic string tension $\sigma_r$ can be extracted from the area-law fall-off of Wilson loops in the $r$-th representation of the color group. I will illustrate the idea on the SU(2) group; its representations are labeled by a ``spin'' index $j$ with half-integer or integer values. Let us assume the vacuum is filled with percolating thin center vortices and divide the loop ${\cal{C}}$ into small patches (with unit area for simplicity). Let $p$ denote the probability that a patch is pierced by a vortex and assume piercings are random and uncorrelated. It is then a simple exercise to show that:
\begin{equation}\label{wloop}
W_j({\cal{C}})\sim\left[p\cdot(-1)^{2j}+(1-p)\cdot(+1)\right]^{{\cal{A}}({\cal{C}})},\qquad {\cal{A}}({\cal{C}})\ \dots\ \mbox{minimal area of the loop }{\cal C}, 
\end{equation}
\begin{equation}\label{sigma_as}
\sigma_j=-\ln\left[1-p+(-1)^{2j}p\right]\approx\left\{\begin{array}{c c l}2p &\dots &\mbox{half-integer }j,\\
0&\dots&\mbox{integer }j.\end{array}\right.
\end{equation}
The asymptotic string tensions are zero for all integer-$j$ representations (i.e.\ those with $N$-ality, or ``biality'', equal to 0), and nonzero and equal for all half-integer $j$ (with $N$-ality equal to 1). The argument can simply be extended to arbitrary $N\ge2$.

An explanation of Casimir scaling at intermediate distances within the center-vortex model was suggested by Faber et al.\ \cite{Faber:1997rp}.\footnote{A related proposal was advanced long ago by Cornwall in a proceedings contribution \cite{Cornwall:1983dx}.} It is attributed to the fact that center vortices in the QCD vacuum do have finite, relatively large thickness.\footnote{Independent ways of estimating the thickness of center vortices are discussed e.g.\ in Section 4 of Ref.\ \cite{Bertle:2000qv}.} If the cross-section of the vortex with the plane of the Wilson loop is fully outside the loop, the vortex does not influence its value. If it is fully inside the loop, the value is multiplied by $(-1)^{2j}$ in SU(2). The effect of a vortex whose core is only partially contained in the area of the loop is modeled by multiplication of the loop by a group element (in the $j$-th representation) that depends on a certain angle $\alpha$ and has random orientation in color space. The angle interpolates between 0 (if the core lies fully outside the loop) and $2\pi$ (if it lies fully inside). This very simple model predicts, as above, (\ref{sigma_as}) for asymptotic distances, and (to a good degree of approximation)
\begin{equation}\label{sigma_intermediate}
\sigma_j\sim j(j+1)\quad\dots\quad\mbox{Casimir scaling}
\end{equation}
for intermediate distances. The model can be generalized to $N\ge2$, at the price of introducing more parameters (angles).\footnote{See Ref.\ \cite{Deldar:1999yy} for particular proposals for SU(3) and SU(4).}

Recently, much attention has been paid to the Yang--Mills theory with the \gtwo\ gauge group~\cite{Holland:2003jy}. The exceptional Lie group \gtwo\ has a trivial center; the center of its universal covering group is also trivial, and there seems to be no reason to believe that the group center plays any important role in \gtwo\ YM theory at all. The theory is at most \textit{temporarily} confining, since potentials for static color charges from any representation, including the fundamental one, must be asymptotically flat. The reason is that any color charge, even a ``quark'' from the (7-dimensional) fundamental representation of \gtwo\ can be screened by a bunch of ``gluons'' (from the 14-dimensional adjoint representation). This fact does not really contradict the center-vortex confinement scenario: \gtwo\ does not possess nontrivial center vortices, and therefore the asymptotic string tension is zero. But even in the \gtwo\ gauge theory, one expects that the static potentials do grow linearly over a certain range of distances, from the scale where perturbation theory breaks to the onset of screening. The linear rise of the fundamental potential in this intermediate region was demonstrated in numerical simulations \cite{Greensite:2006sm}.

Dimensional reduction can be invoked to argue that even in \gtwo\ YM theory one would expect Casimir scaling of potentials at intermediate distances. But if, in the language of field configurations,  finite thickness of center vortices was responsible for Casimir scaling in SU($N$), what should it be attributed to in a theory with trivial center? A common origin of Casimir scaling for ``center-poor'' and ``center-rich'' gauge models was suggested in Ref.\ \cite{Greensite:2006sm}. We extended and improved the model of Ref.\ \cite{Faber:1997rp}. We assume the Yang--Mills vacuum state has a domain structure, with the color magnetic flux in each domain quantized in units of the gauge group center (be it trivial or not), and Casimir scaling results from random spatial fluctuations of the flux in each domain.

In this model, it is assumed that if we take a 2D slice of the four-dimensional volume, we can split it into domains (``patches'') of a typi\-cal area $A_d$. Within each domain color magnetic fields fluctuate randomly and (almost) independently, with a short length of correlation $l$. Each domain contains small independently fluctuating subregions of area $l^2\ll A_d$. The only constraint on fluctuating fields bounds the total magnetic flux over each domain to correspond to an element of the gauge-group center. In SU(2), there are domains of the center-vortex type and of the vacuum type. The former correspond to the nontrivial center element, $-\mathbf{I}$, and represent a cross section of a thick center vortex; the latter carry a zero total magnetic flux. In \gtwo\ all domains will be of the vacuum type, since the center contains the identity element only.
	
I will only summarize conclusions of the model: If the center of the gauge group contains $N$ elements, there are $N$ types of domains enumerated by the value $k\in \{0, 1, \dots, N-1\}$, each assumed to appear with a probability $f_k$ centered at any given plaquette in the plane of the loop. The effect of a domain (a 2D cross section of the $k$th vortex) on a planar Wilson loop is to multiply the loop by a group element
\begin{equation}
\mathcal{G}(\alpha^{(k)},\mathcal{S})=\mathcal{S}\;\exp\left[i\vec{\alpha}^{(k)}\cdot\vec{\mathcal{H}}\right]\; \mathcal{S}^\dagger,
\end{equation}
where $\{\mathcal{H}_i\}$ are generators of the Cartan subalgebra, $\mathcal{S}$ is a random element of the group, 
and angles $\vec{\alpha}^{(k)}$ depend on the location of the vortex/domain with respect to the loop. If the domain is all contained within the loop area, then
\begin{equation}
\exp\left[i\vec{\alpha}^{(k)}\cdot\vec{\mathcal{H}}\right]=z_k\mathbf{I},
\end{equation}
where $z_k$ is the $k$th center element, $\mathbf{I}$ is the unit element. If the domain is outside the loop, it has no effect, i.e.
\begin{equation}
\exp\left[i\vec{\alpha}^{(k)}\cdot\vec{\mathcal{H}}\right]=\mathbf{I}.
\end{equation}
For a Wilson loop from the representation $r$, the averaged contribution of a domain is
\begin{equation}
\mathcal{G}_r(\alpha^{(k)})\; \mathbf{I}_{d_r}=\frac{1}{d_r}\chi_r\left(\exp\left[i\vec{\alpha}^{(k)}\cdot\vec{\mathcal{H}}\right]\right)\; \mathbf{I}_{d_r},
\end{equation}
where $d_r$ is the dimension of the representation $r$ and $\mathbf{I}_{d_r}$ is the $d_r\times d_r$ unit matrix.

One further assumes that probabilities to find domains of any type centered at two different plaquettes are independent, and that for loops smaller than the typi\-cal size of the domain, the r.m.s.\ of phases $\alpha$ is proportional to the area of the vortex contained in the interior of the loop. Then, both in SU($N$) and \gtwo, the static potential $V_r(R)$ of the representation $r$ will be linearly rising for distances $l\ll R\ll\sqrt{A_d}$, with a string tension approximately proportional to its quadratic Casimir:
\begin{equation}
\sigma_r\sim C_r\quad\dots\quad\mbox{intermediate distances.}
\end{equation}
For very large Wilson loops most vortices will be contained within the loop, the average phases are proportional to the corresponding total magnetic flux through the domain, and the prediction is, for $R\gg\sqrt{A_d}$,
\begin{equation}
\sigma_r\sim F[k_r, \{f_k\}]\quad\dots\quad\mbox{asymptotic distances}
\end{equation}
for SU($N$), where $k_r$ is the $N$-ality of the representation~$r$, and 
\begin{equation}
\sigma_r=0\quad\dots\quad\mbox{asymptotic distances}
\end{equation}
for the \gtwo\ gauge group.

\section{Casimir scaling in G${}_2$ gauge theory}\label{g2_cs}
The described domain model of the YM vacuum with magnetic disorder, containing only a few adjustable parameters, predicts Casimir scaling of higher representation potentials at intermediate distances not only for SU($N$), but also for \gtwo\ gauge theory. The prediction can be verified or disproved in numerical simulations of the theory in lattice formulation. This is in principle a straightforward task, but not cheap for \gtwo: simulations are quite demanding on computer resources, determination of string tensions requires all machinery developed in the past for calculating potentials (anisotropic lattices, ground-state overlap enhancement, smearing) plus some bits and pieces of information from group theory. Technical details of our calculations can be found in Ref.\ \cite{Liptak:2008gx}, I will present here only a subset of representative results. 

We simulated \gtwo\ lattice gauge theory with the Wilson action on anisotropic $L^3\times(2L)$ lattice:
\begin{equation}
\displaystyle S = - \frac{\beta}{7} \left\{ \xi_0 \sum_{x, i > 0}\mbox{Re Tr}\left[   P_{i0} (x)  \right]  
- \frac{1}{\xi_0} \sum_{x, i > j > 0} \mbox{Re Tr}\left[   P_{ij} (x)  \right]\right\}
\end{equation}
with the bare-anisotropy parameter $\xi_0$ tuned so that the physical anisotropy $\xi=a_s^\mathrm{phys}/a_t^\mathrm{phys}$ equals~2. To increase overlap of the trial quark-antiquark state with the ground state, we used in the computation of Wilson loops spatial links smeared by a \gtwo\ generalization of the stout smearing method~\cite{Morningstar:2003gk}. Most of our results come from $14^3\times 28$ lattice at three values of the coupling $\beta=9.5, 9.6, 9.7$ on the weak-coupling side of the crossover region observed in Ref.\ \cite{Holland:2003jy}.

The potential between a static quark and antiquark from the representation $\{D\}$ of the \gtwo\ group can be determined from values of Wilson loops $W_{\{D\}}(r,t)$ by a fit of the form
\begin{equation}
-\ln W_{\lbrace D\rbrace}(r,t)=
C_{\lbrace D\rbrace}+V_{\lbrace D\rbrace}(r)\cdot t
\end{equation}
in an appropriate interval $(t_\mathrm{min},t_\mathrm{max})$. The resulting potentials are then parametrized by the usual Coulomb plus linear form:
\begin{equation}\label{fit}
V_{\lbrace D\rbrace}(r)=c_{\lbrace D\rbrace}-\frac{\alpha_{\lbrace D\rbrace}}{r}+\sigma_{\lbrace D\rbrace}\;r.
\end{equation}

\gtwo\ irreducible representations $\{D\}$ are labeled by two Dynkin coefficients $[\lambda_1,\lambda_2]$, the dimension of the representation is given by \cite{Engelfield:1980ab}:
\begin{equation}
D=\frac{9}{40}(\ell_1^2 - \ell_2^2)(\ell_2^2 - \ell_3^2)(\ell_3^2 - \ell_1^2),
\end{equation}
where $\ell_1 = \frac{1}{3} (1+\lambda_1)$, $\ell_2 = \frac{1}{3} (4 + \lambda_1 + 3\lambda_2)$, and $\ell_3 = \frac{1}{3} (5+2\lambda_1 + 3\lambda_2)$. The ratio of eigenvalues of the quadratic Casimir operator is \cite{Engelfield:1980ab}:
\begin{equation}
d_{\{D\}}\equiv{\displaystyle \frac{C_{\{D\}}}{C_F}}=\frac{1}{4}\left(\ell_1^2 + \ell_2^2 + \ell_3^2 - \frac{14}{3}\right),
\end{equation}
where $C_F$ is the quadratic Casimir for the fundamental representation. 

\begin{figure}[b!]
\centering
\begin{tabular}{c c}
\includegraphics[width=0.45\textwidth]{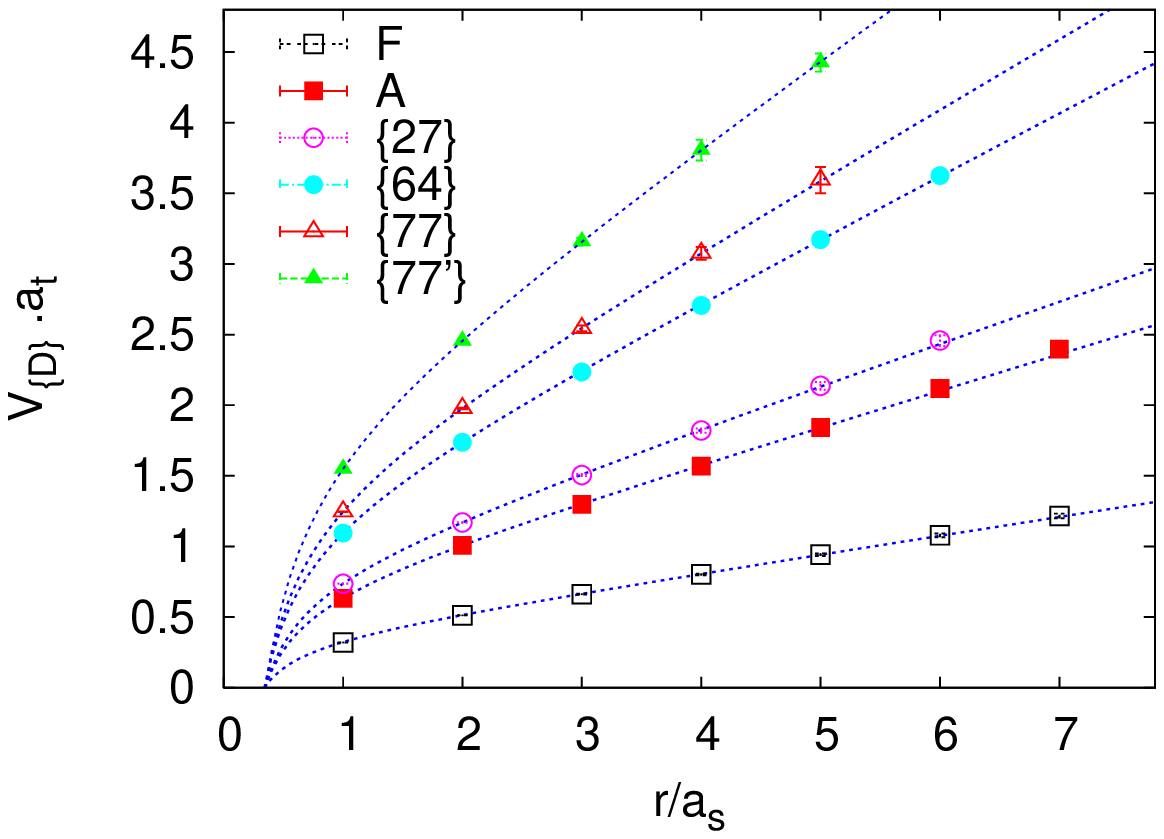}&
\includegraphics[width=0.45\textwidth]{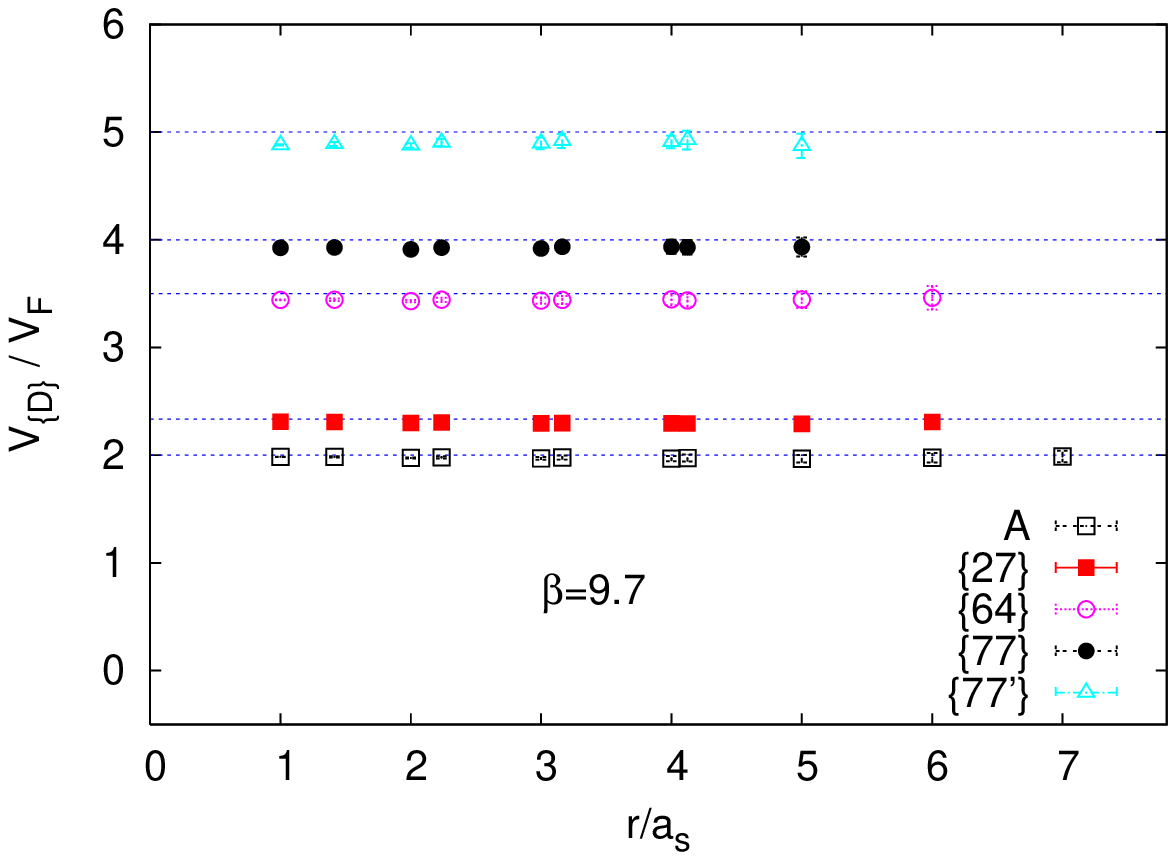}
\end{tabular}
\caption{[Left:] Potentials for different representations vs.\ the dimensionless $r$; $14^3 \times 28$ lattice, $ \beta = 9.6$. [Right:] Ratios ${V}_{\{D\}} / {V}_{F}$ for different representations $\{D\}$ as functions of the dimensionless $r$; $14^3 \times 28$ lattice, $\beta=9.7$. Horizontal lines show Casimir-scaling predictions.}\label{g2pot}
\end{figure}
\begin{table}[t!]
\centering
\begin{tabular}[c]{ c  c  c c  c c } 
\hline
 $\beta$ &  $A$ &  $\{27\}$ & $\{64\}$ & $\{77\}$ &  $\{77 ' \}$   \\\hline\hline
 9.5\phantom{${}^\textrm{1}$}         & 1.88(4) & 2.15(5) &  3.1(1) &    ---       & --- \\
 9.6\phantom{${}^\textrm{1}$}         & 1.94(4) & 2.24(6) &  3.35(8)   & 3.8(2)  & 4.6(2) \\
 9.7\phantom{${}^\textrm{1}$}         & 1.96(6) & 2.28(7) &  3.5(1) & 4.0(2) &  4.9(2) \\ \hline
{CS}\phantom{${}^\textrm{1}$} &  {2.0}      &  {2.333}      &  {3.5}      &  {4.0}      &  {5.0}     \\\hline
\end{tabular}
\caption{Ratios $\sigma_{\lbrace D\rbrace}/\sigma_F$ for the adjoint (labeled $A$), $\{27\}$, $\{64\}$, $\{77\}$, and $\{77 ' \}$ representations. The last line is the Casimir-scaling prediction.}\label{ratios}
\end{table}

\begin{figure}[t!]
\centering
\begin{tabular}{c c}
\includegraphics[width=0.45\textwidth]{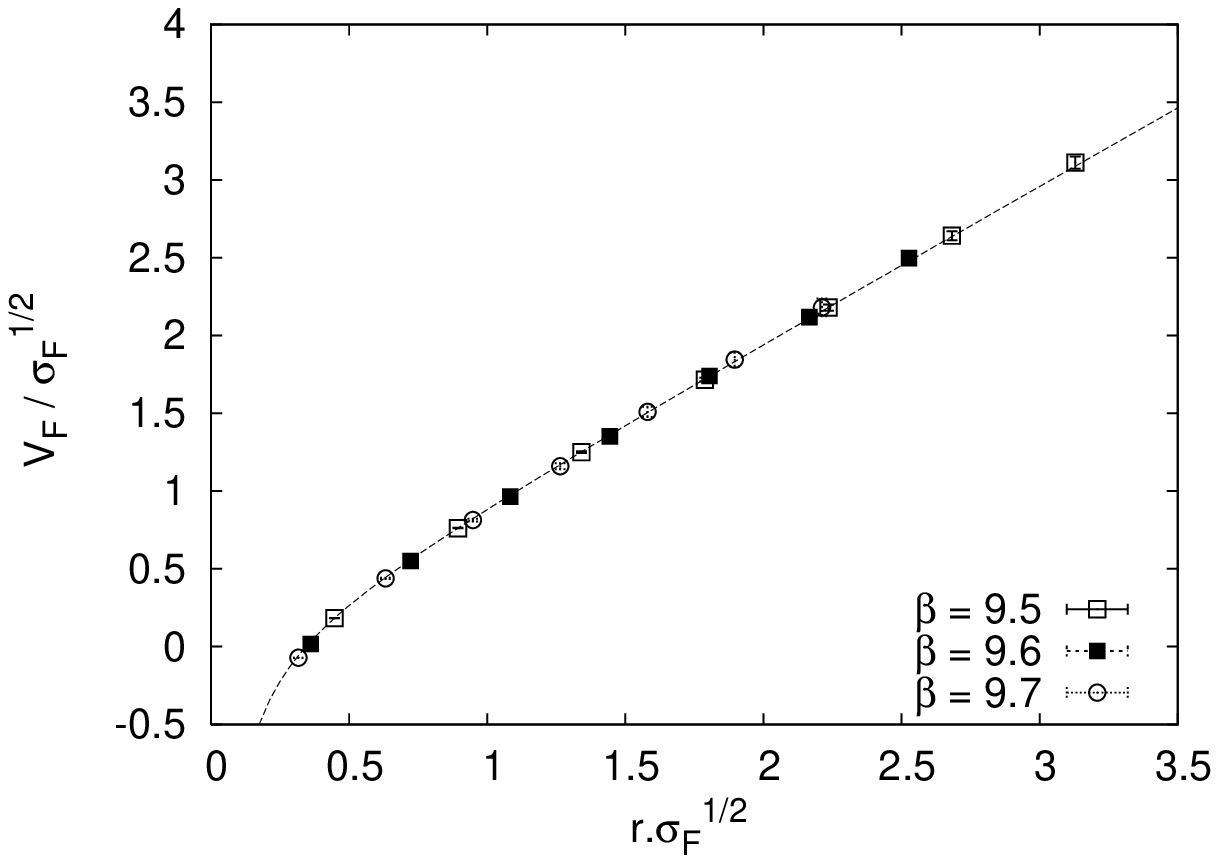}&
\includegraphics[width=0.45\textwidth]{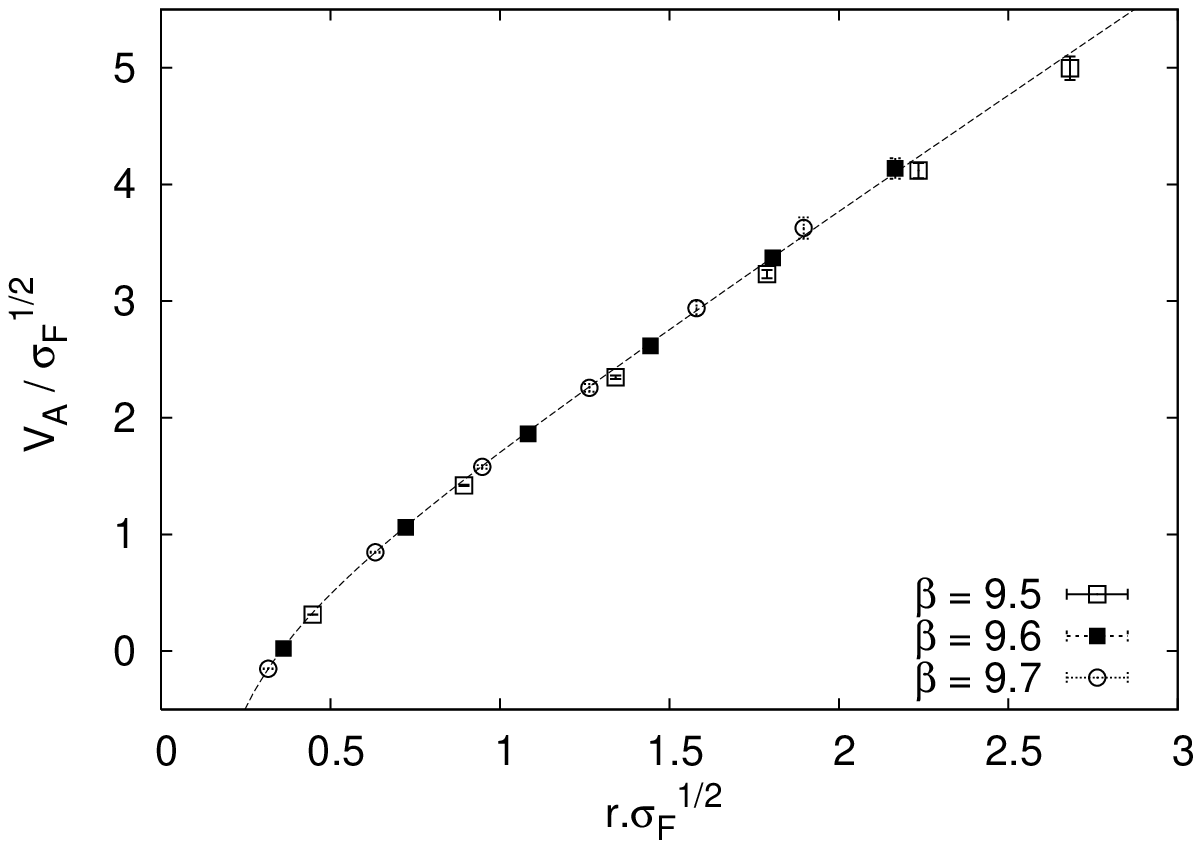}
\end{tabular}
\caption{The fundamental (left) and adjoint (right) representation potential in physical units, all three $\beta$ values, from $14^3\times 28$ lattice. (The constant $c$, cf.\ Eq.~(\protect\ref{fit}), is subtracted.)}\label{g2pot_phys}
\end{figure}

Results for string-tension ratios for color sources from five \gtwo\ representations are summarized in Table \ref{ratios}, potentials are displayed in Fig.\ \ref{g2pot} for $\beta=9.6$ and $9.7$. Fundamental and adjoint potentials for all three couplings $\beta$ are shown together in Fig.\ \ref{g2pot_phys}, expressed in physical units defined by the fundamental string tension. Our results convincingly demonstrate (approximate) Casimir scaling for static potentials between color charges from various representations of \gtwo.\footnote{Casimir scaling of \gtwo\ potentials has recently been seen also in numerical simulations of \gtwo\ lattice gauge theory in $(2+1)$ dimensions \cite{Wellegehausen:2009rq}.} The agreement between measured values of intermediate string tensions with predictions based on values of quadratic Casimirs is quite striking; they differ by at most 10--15\%, and this can hardly be just a numerical coincidence. The results of course cannot prove that the model described in Section \ref{model} is right, but combined with the solid evidence for Casimir scaling in SU(2) and SU(3), they provide support for its main ingredients (common for all groups) -- a magnetically disordered vacuum with a domain structure.

\section{A suggestion for an approximate vacuum wave-functional of Yang--Mills theory in $(2+1)$ dimensions}\label{wf}
Can one derive (at least some) elements of the described picture from first principles? Recently, we have suggested an approximate form of the vacuum wave-functional which solves the SU(2) Yang-Mills Schr\"odinger equation in the temporal gauge \cite{Greensite:2007ij}:\footnote{Expressions below are all assumed to be properly defined on a lattice, with lattice spacing serving as regulator, but for simplicity I will mostly use continuum notations.}
\begin{equation}\label{ourWF}
\Psi_0[A]{=}
\exp\left[-\frac{1}{2}\displaystyle\int d^2xd^2y\; 
B^a(x){\displaystyle
\left(\frac{1}{\sqrt{-{\cal D}^2-
\lambda_0+m^2}}\right)_{xy}^{ab}}B^b(y)\right].
\end{equation}
Here ${B^a(x)}=F_{12}^a(x)$ is the color magnetic field strength, ${\cal D}^2$ the covariant laplacian in the adjoint representation:
\begin{equation}
\left({-{\cal D}^2}\right)^{ab}_{xy}=
\displaystyle\sum_{k=1}^2 \left[2\delta^{ab}\delta_{xy}-
{\cal U}^{ab}_k(x)\delta_{y,x+\hat{k}}-{\cal U}^{\dagger ba}_k(x-\hat{k})\delta_{y,x-\hat{k}}\right],
\end{equation}
where ${\cal U}_k(x)$ are the link fields in the adjoint representation, ${\lambda_0}$ denotes the lowest eigenvalue of $(-{\cal D}^2)$, and ${m}$ is a constant (mass) proportional to $g^2\sim 1/\beta$.

The proposed wave-functional (\ref{ourWF}) is reminiscent of Samuel's \cite{Samuel:1996bt}, the difference is that in his proposal a single free parameter $m^2_0$ replaces our $(-\lambda_0+m^2)$ in the denominator. The reason for subtracting the lowest eigenvalue from the operator $(-{\cal D}^2)$ is that our numerical simulations indicate that its spectrum  might diverge in the continuum limit.

This wave-functional has quite a few attractive properties:
\begin{enumerate}
\item
In the free-field limit ($g\to 0$), the covariant laplacian turns into ordinary laplacian, $\lambda_0$ and $m$ go to 0, and $\Psi_0[A]$ becomes the well-known vacuum wave-functional of electrodynamics: 
\begin{equation}
\Psi_0[A]\;
{=}\;
\exp\left\{-\frac{1}{2}\displaystyle\int d^2xd^2y\; 
[\nabla\times A^a(x)]{\displaystyle
\left(\frac{\delta^{ab}}{\sqrt{-\nabla^2}}\right)_{xy}}[\nabla\times A^b(y)]\right\rbrace.
\end{equation}
\item
Eq.\ (\ref{ourWF}) is a good approximation to the true vacuum also in completely different corner of the configuration space, namely if we restrict to fields constant in space and varying only in time. In the large-volume limit the solution of the YM Schr\"odinger equation in that case is, up to $1/V$ corrections:
\begin{equation}
\Psi_0=\exp[-VR_0]=\exp\left[\displaystyle
-\frac{1}{2}gV\frac{(\vec{A_1}\times\vec{A_2})\cdot(\vec{A_1}\times\vec{A_2})}
{\sqrt{\vec{A}_1\cdot\vec{A}_1+\vec{A}_2\cdot\vec{A}_2}}
\right],
\end{equation}
and exactly the same expression follows from (\ref{ourWF}) assuming $\vert g \vec{A}_{1,2}\vert \gg m, \sqrt{\lambda_0}$.
\item
If we divide the field strength $B^a(x)$ into ``fast'' and ``slow'' components, the portion of the (squared) vacuum wave-functional in $B_\mathrm{slow}$ is approximately 
\begin{equation}\label{dim_red}
\vert\Psi_0\vert^2\approx\displaystyle\exp\left[
-\frac{1}{m}\int d^2x\; 
B^a_\mathrm{slow}(x)\;B^a_\mathrm{slow}(x)\right].
\end{equation}
This is exactly the probability measure for YM theory in two Euclidean dimensions, which (i)~is confining for $m>0$, and (ii) exhibits Casimir scaling for string tensions of all color-charge representations. The fundamental string tension is easily computed as $\sigma_F(\beta)=3 m/(4\beta)$. The last expression can be used to fix the value of $m$ in Eq.\ (\ref{ourWF}) at a given $\beta$ from the known value of $\sigma_F$. 
\item
Confinement requires $m$ to be larger than zero; if one takes $m$ in the wave-functional (\ref{ourWF}) as a variational parameter and computes (approximately) the expectation value of the Yang--Mills Hamiltonian, one finds that non-zero (and finite) value of $m$ is energetically preferred.
\item
If we fix the mass $m$ in the wave-functional to get the right string tension $\sigma_F$ at a given $\beta$, we can test our proposal by calculating e.g.\ the mass gap of the theory. We have proposed a recursive procedure for generating independent lattice configurations with the probability distribution given by the square of the wave-functional (\ref{ourWF}), the interested reader should consult Ref.\ \cite{Greensite:2007ij} for its detailed exposition. We call two-dimensional lattice configurations obtained in this way ``recursion lattices''. One can compute various observables with these lattices, and compare results with ``MC lattices'', i.e.\ two-dimensional slices of lattices generated in a full $D=3$ lattice Monte Carlo simulation. To extract the mass gap, one computes
the correlator
\begin{equation}
{\cal G}(x-y)=\left\langle (B^aB^a)_x
(B^b B^b)_y\rangle-
\langle (B^aB^a)_x\right\rangle^2,
\end{equation}
and determines the mass gap $M$ by fitting the result to the form
\begin{equation}
\mbox{(const.)}\left(1+{\textstyle\frac{1}{2}}M R\right)^2\frac{e^{-M R}}{R^6},\qquad R=\vert x-y\vert.
\end{equation}
Fig.\ \ref{gap} shows our values for the mass gap versus $\beta$ extracted from recursion-lattice data. It turns out that, given the asymptotic string tension as input, we can compute the mass gap fairly accurately from our wave-functional. The discrepancy from Monte Carlo results of Meyer and Teper \cite{Meyer:2003wx} for the $0^+$ glueball mass are at the level of at most a few ($<6$) percent.
\begin{figure}[t!]
\centerline{\includegraphics[width=0.5\textwidth]{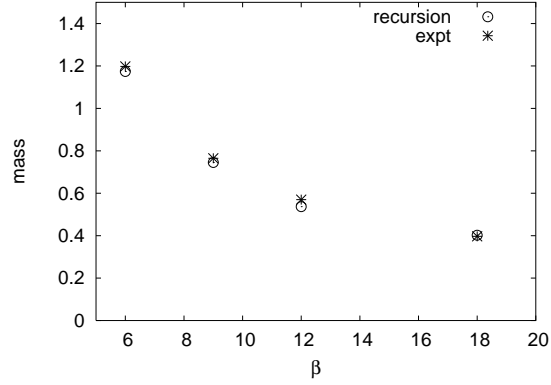}}
\caption{Mass gaps extracted from recursion lattices at various lattice couplings, compared to $0^+$ glueball masses in 2 dimensions obtained in Ref.\ \cite{Meyer:2003wx} (denoted ``expt'') via standard lattice Monte Carlo methods.}
\label{gap}
\end{figure}
\item
The YM wave-functional in Coulomb gauge can be obtained by restricting the temporal-gauge wave-functional to transverse gauge fields. We have computed the color-Coulomb potential and the Coulomb-gauge ghost propagator using recursion and MC lattices and the results are in reasonable agreement, see Jeff Greensite's talk at this workshop \cite{Greensite:2009aa}.
\end{enumerate}

The dimensional reduction form (\ref{dim_red}) at large distances implies an area law fall-off for large Wilson loops, and also Casimir scaling of higher-representation Wilson loops. The question is how Casimir scaling turns into $N$-ality dependence, how color screening enters the game within this setting. It may be necessary to introduce into our wave-functional additional term(s), e.g.\ a gauge-invariant gluon-mass term advocated by Cornwall \cite{Cornwall:2007dh}. However, there are indications that terms needed for color screening might be contained in our simple wave-functional and would appear as corrections to the dimensional-reduction form (\ref{dim_red}).

Let us write the lattice vacuum state $\Psi_0[U]$ in the form $\exp(R[U])$. Greensite \cite{Greensite:1979yn} developed a strong-coupling technique for calculating $R[U]$ in Hamiltonian lattice gauge theory. $R[U]$ is expressed as a sum over spacelike Wilson loops and products of loops on the lattice, schematically:

\begin{equation}
\centerline{\includegraphics[width=0.5\textwidth,clip,trim=0 0 0 30mm]{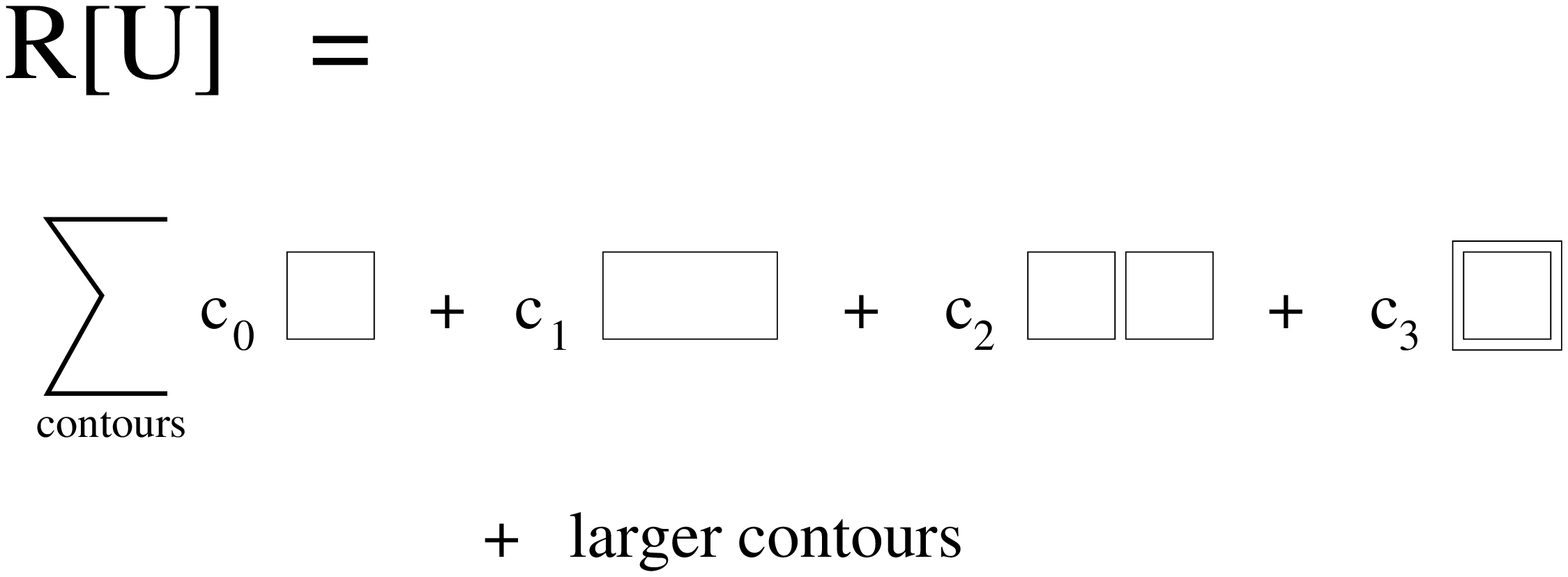}}
\end{equation}

\begin{figure}[h!]
\centerline{\includegraphics[angle=90,width=0.4\textwidth]{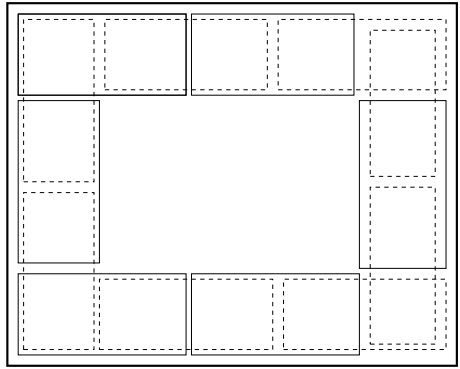}}
\caption{Tiling/screening an adjoint
Wilson loop by $1\times2$ rectangles. The loop is denoted by a heavy solid line.}
\label{screening}
\end{figure}

\noindent The first few coefficients $c_i$ for SU(2) lattice gauge theory in $(2+1)$ dimensions were computed in Ref.\ \cite{Guo:1994vq} and for smoothly varying fields one gets:
\begin{equation}\label{guo}
{\displaystyle\Psi_0[U]=\exp\left[
-\frac{2}{\beta}\int d^2x\;\left(a\kappa_0B^2-a^3
\kappa_2B(-{\cal D}^2)B+\dots\right)\right]},
\end{equation}
where $a$ is the lattice spacing, $\kappa_0=\frac{1}{2}c_0+2(c_1+c_2+c_3)$, $\kappa_2=\frac{1}{4}c_1$, $c_0$ is ${\cal O}(\beta^2)$, and $c_1, c_2, c_3$ are ${\cal O}(\beta^4)$. The leading correction to dimensional-reduction form is contained in the term proportional to $\kappa_2$, and comes from the $1\times2$ loop in $R[U]$ proportional
to $c_1$. If we evaluate the adjoint Wilson loop $W_A({\cal C})$ in the strong-coupling expansion, the $c_1$ term provides the leading perimeter-law contribution, sketched 
in Fig.\ \ref{screening}, proportional to $\left({c_1}/{2}\right)^{{\cal{P}}({\cal{C}})-4}$. So the term that gives the leading correction to the dimensional-reduction form is also responsible for screening of adjoint loops.

If we now work out the leading correction to the dimensional-reduction form (\ref{dim_red}) by expanding the kernel in our wave-functional (\ref{ourWF}), we get:
\begin{equation}\label{correction}
\vert\Psi_0\vert^2=\displaystyle\exp\left[-\frac{1}{m}\int d^2x\;\left(B_\mathrm{slow}B_\mathrm{slow}-B_\mathrm{slow}\frac{-{\cal D}^2-\lambda_0}{2m^2}B_\mathrm{slow}+\dots\right)\right].
\end{equation}
One should note a striking similarity of the second term in Eq.\ (\ref{correction}) to the $\kappa_2$-term in Eq.\ (\ref{guo}). This gives some hope that the proposed approximate wave-functional (\ref{ourWF}) may in fact not only incorporate Casimir scaling of string tensions at intermediate distances, but also $N$-ality dependence in the asymptotic region. 

\section{Conclusions}\label{conclusions}

\begin{itemize}
\item
Casimir scaling is a natural outcome of a model of the QCD vacuum as a medium in which color magnetic fields fluctuate almost independently and are only weakly constrained by the condition that the total flux through a cross-section of the domain corresponds to an element of the gauge group center.
\item
Lattice data show that string tensions of higher-representation potentials at intermediate distances satisfy Casimir scaling to surprising accuracy for SU(2), SU(3), and even \gtwo\ gauge theory.
\item
Some elements of the model are (or may be) contained in the simple approximate form of the confining Yang--Mills vacuum wave-functional in $(2+1)$ dimensions proposed in Ref.~\cite{Greensite:2007ij}.
\end{itemize}

\acknowledgments{I thank Jeff Greensite, Kurt Langfeld, {L\kern-.08cm'}udov{\'\i}t Lipt\'ak, Hugo Reinhardt, and Torsten Tok for collaboration on topics covered in the present contribution. I am grateful to Jeff Greensite for comments to the manuscript. I would also like to express gratitude to Mike Corn\-wall for the invitation to the interesting and stimulating workshop in Trento.}

\end{document}